\documentclass[12pt,prd,draft,nofootinbib]{revtex4}
\usepackage{amsfonts}
\usepackage{amssymb}
\usepackage{amsmath}
\usepackage{bm}
\usepackage{latexsym}
\newcommand{\beq}{\begin{eqnarray}}
\newcommand{\eeq}{\end{eqnarray}}
\newcommand{\be}{\begin{equation}}
\newcommand{\ee}{\end{equation}}
\def\la{\mathrel{\mathpalette\fun <}}

\def\fun#1#2{\lower3.6pt\vbox{\baselineskip0pt\lineskip.9pt
\ialign{$\mathsurround=0pt#1\hfil ##\hfil$\crcr#2\crcr\sim\crcr}}}

\newcommand{{\SD}}{\rm SD}

\newcommand{\vep}{\bm p}

\newcommand{\veL}{\bm L}

\newcommand{\veS}{\bm S}

\newcommand{\lan}{\langle}
\newcommand{\ran}{\rangle}

\begin{document}

\title{Dominant spin-orbit effects in radiative decays $\Upsilon(3S)\rightarrow \gamma+\chi_{bJ}(1P)$}

\author{\firstname{A.M.}~\surname{Badalian}}
\email{badalian@itep.ru}
\affiliation{State Research Center, Institute of Theoretical and
Experimental Physics, Moscow, 117218 Russia}

\author{\firstname{B.L.G.}~\surname{Bakker}}
\email{b.l.g.bakker@vu.nl}
\affiliation{Department of Physics and Astronomy,
Vrije Universiteit, Amsterdam, The Netherlands}

\date{\today}

\begin{abstract}
We show that there are two reasons why the partial width
for the transition $\Gamma_1(\Upsilon(3S)\rightarrow
\gamma+\chi_{b1}(1P))$ is suppressed. First, the spin-averaged
matrix element $\overline{I(3S|r|1P_J)}$ is small, being equal to
0.023~GeV$^{-1}$ in our relativistic calculations. Secondly, the
spin-orbit splittings produce relatively large contributions, giving
$I(3S|r|1P_2)=0.066$~GeV$^{-1}$, while due to a large cancellation the
matrix element  $I(3S|r|1P_1)=-0.020$~GeV$^{-1}$ is small and negative;
at the same time the magnitude of $I(3S|r|1P_0)=-0.063$~GeV$^{-1}$
is relatively large. These  matrix elements  give rise to the
following partial widths: $\Gamma_2(\Upsilon(3S)\rightarrow
\gamma+\chi_{b2}(1P))=212$~eV, $\Gamma_0(\Upsilon(3S)\rightarrow
\gamma+\chi_{b0}(1P))=54$~eV, which are in good agreement with the
CLEO and BaBar data, and also to $\Gamma_1(\Upsilon(3S)\rightarrow
\gamma+\chi_{b1}(1P))=13$~eV, which satisfies the BaBar limit,
$\Gamma_1({\rm exp}.) < 22$~eV.
\end{abstract}

\maketitle

\section{Introduction}

In recent years, several new bottomonium states were discovered due to
studies of radiative decays \cite{ref.1,ref.2,ref.3,ref.4}. In \cite{ref.1}
CLEO has observed the $\Upsilon(1D)$ in the four-photon decay cascade,
$\Upsilon(3S) \rightarrow \gamma+\chi_b(2P),~~\chi_b(2P)\rightarrow
\gamma+\Upsilon(1D),~~\Upsilon(1D)\rightarrow \gamma+\chi_b(1P),~~
\chi_b(1P)\rightarrow  \gamma+\Upsilon(1S)$, and later this state was
observed by BaBar in another four-photon cascade via the $\Upsilon(2S)$
\cite{ref.2}. In 2008 a new state, $\eta_b(1P)$, was discovered
by BaBar, first in radiative decay $\Upsilon(3S)\rightarrow
\gamma+\eta_b(1S)$ \cite{ref.3} and then in $\Upsilon(2S)\rightarrow
\gamma+\eta_b(1S)$ \cite{ref.4}; later $\eta_b(1S)$ was confirmed
by CLEO \cite{ref.5}. Moreover, new or more precise data on
different radiative transitions, like $\Upsilon(3S)\rightarrow
\gamma+\chi_b(n\,{}^3P_J)~(n=1,2),~~ \chi_b(1P,2P)\rightarrow
\gamma+\Upsilon(1S)$, and $\chi_b(2P)\rightarrow \gamma+\Upsilon(2S)$,
were presented in Refs.~\cite{ref.6,ref.7,ref.8,ref.9}.

This new experimental information is of a special importance for the
theory to provide a better understanding of the role of relativistic and
spin-dependent effects in bottomonium, and may be used as a test of
different models and approximations. There are a large number of papers
devoted to radiative decays in bottomonium
\cite{ref.10,ref.11,ref.12,ref.13,ref.14},
and a comparison of different results was already presented in
\cite{ref.12,ref.13,ref.14}, where the predicted partial widths are
shown to be rather close to each other for most radiative $\rm E1$
transitions and to agree with the existing experimental data. The
only exception is the radiative decays $\Upsilon(3S)\rightarrow
\gamma+\chi_b(1P_J)~(J=0,1,2)$, which are discussed in detail in
\cite{ref.14}. Their partial widths are defined by the matrix element
(m.e.) $I(3S|r|1P_J)\equiv \lan\Upsilon(3S)|r|1\,{}^3P_J\ran~~(J=0,1,2)$
and below we shall also use the spin-averaged m.e., denoted as
$\overline{I(3S|r|1P)}$.

These m.e. strongly differ in the nonrelativistic (NR) and relativistic
cases, even within the same model. The predicted transition rate
$\Gamma_1(\Upsilon(3S)\rightarrow \gamma+\chi_{b1}(1P))$ varies
in a wide range, $(3-110)$~eV \cite{ref.14} and is in many cases
larger than the experimental width: $\Gamma_1=(33\pm 10)$~eV from
the CLEO data \cite{ref.8}); a smaller value $\Gamma_1=(10^{+8}_{-6})$~eV
was measured by BaBar \cite{ref.9}. Moreover, even in the models
which predict a small partial width $\Gamma_1$, their other two
rates, $\Gamma_J(J=0,2)$, do not agree with the experimental
values \cite{ref.15}. Therefore, the ratio of the transition
rates, $r_{1,0}=\frac{\Gamma_1(\Upsilon(3S)\rightarrow
\gamma+\chi_{b1}(1P))}{\Gamma_0(\Upsilon(3S)\rightarrow
\gamma+\chi_{b0}(1P))}$, must be considered an important characteristic,
which is small in experiments: $r_{1,0}\sim 0.5$ from the CLEO
\cite{ref.8} and $r_{1,0}\sim 0.2$ from the BaBar data \cite{ref.9}.

The m.e. $I(3S|r|1P_J)$ may differ several times in NR and
relativistic calculations, even within the same model or while
different static potentials are used \cite{ref.11,ref.16}. In
Ref.~\cite{ref.17} the suppression of this m.e. was shown to be
quite strong in the NR limit for the power-law potentials
$V(r)\sim r^{\alpha}$ with $-1 <\alpha <2$. Since in bottomonium,
even for $\Upsilon(3S)$, the relativistic corrections are not
large, $\frac{\vep^2}{m_b^2}\la 0.1$, one may assume that this
fact occurs because of the different asymptotics of the wave
functions (w.f.) of the Schr\"{o}dinger and relativistic
equations.

An interesting result was obtained in Ref.~\cite{ref.16}, where
for the  NR Hamiltonian the partial width
$\Gamma_2=\Gamma(\Upsilon(3S)\to\gamma+\chi_{b2}(1P))$ decreases
ten times, if instead of the Cornell potential with $\alpha({\rm
static})=$~constant, the Wisconsin potential which takes into
account the asymptotic freedom behavior of the vector strong
coupling, is used. This result reminds of the situation with the
dielectron widths of $\Upsilon(nS)~(n=1,2,3)$, where agreement
with experiment is reached only for the potential with the
asymptotic freedom behavior of the strong coupling \cite{ref.18}.

However, even for this kind of potentials the spin-averaged m.e.
$\overline{I(3S|r|1P)}$ appears to depend on the freezing
(critical) value of the vector strong coupling used. In this paper
we consider gluon-exchange (GE) potentials with two different values
of $\alpha_{\rm crit}$.

It is also evident that since the m.e. $\overline{I(3S|r|1P_J)}$
is small, it may strongly depend on other small effects, in
particular, on the spin-orbit interaction used. Here we show that
due to the spin-orbit splittings the m.e. $I(3S|r|1P_J)$ acquire
corrections of the same order as the value of the spin-averaged
m.e. $\overline{I(3S|r|1P)}$, and a large cancellation takes place
in the m.e. with $J=1$. Here in our calculations we use the
relativistic string Hamiltonian (RSH) \cite{ref.19}, which was
already tested in a number of papers, devoted to different
bottomonium properties \cite{ref.20}.

\section{Radiative decays}

Electric dipole transitions between an initial state $i = 3\,{}^3S_1$,
and a final state $f = 1\,{}^3P_J$, are defined by
the partial width \cite{ref.10,ref.11,ref.12,ref.13, ref.14},
\begin{equation}
\Gamma(\,\,i  \stackrel{\mathrm{E1}}{\longrightarrow} \gamma +
f\,\,) = \frac{4}{3}\,\alpha\,
e_{Q}^{2}\,E_\gamma^{3}\,(2J^{\prime}+1)\,{\rm S}^{\rm E}_{if}
\,|\mathcal{E}_{if}|^{2}~~,~ \label{eq.1}
\end{equation}
where $J^{\prime}=J_f,~l^{\prime}=l_f$, and the statistical factor
${\rm S}^{\rm E}_{if}={\rm S}^{\rm E}_{fi}$ is given by
\begin{equation}
{\rm S}^{\rm E}_{if} = \max{(l,l^{\prime})} \left\{
          \begin{array}{ccc}
            J & 1 & J^{\prime}  \\
            \l^{\prime} & s & l
            \end{array}\right\}^{2}~~~\label{eq.2}
\end{equation}
and for the transitions between the $n\,{}^3S_1$ and $m{~}^3P_J$
states with the same spin $S=1$ this coefficient $S_{if}^{E}=1/9$.

The RSH is simplified in the case of bottomonium, where in the
Hamiltonian the string and self-energy corrections can be neglected
because they are very small, $\leq 1$~MeV. Then the original form of
the RSH with the static potential
\begin{equation}
V_{\rm B}(r)=\sigma r - \frac{4}{3} \frac{\alpha_{\rm B}(r)}{r}
\label{eq.5}
\end{equation}
is
\begin{equation}
 H =\frac{{\vep}^2+m_b^2}{\omega}+\omega+V_B(r).
\label{eq.3}
\end{equation}
Here $m_b$ is the $b$-quark pole mass, while the value of $\omega$ is
determined from the extremum condition $\frac{\partial
H}{\partial \omega}=0$, which gives $\omega= \sqrt{\vep^2+m_b^2}$,
being equal to the kinetic energy of a $b$ quark. Substituting this $\omega$
into Eq.~(\ref{eq.3}) one arrives at the spinless Salpeter equation
(SSE):
\begin{equation}
 H_0=2\sqrt{\vep^2+m_b^2} + V_{\rm B}(r),
\label{eq.4}
\end{equation}
The kinetic term occurring in (\ref{eq.4}) is widely used in relativistic
potential models \cite{ref.21,ref.22,ref.23}, however, as compared to
constituent potential models, the RSH has several important differences.

\begin{enumerate}

\item By derivation, the mass of the $b$ quark in the kinetic term
cannot be chosen arbitrarily: it must be equal to the pole mass of
a $b$ quark, which takes into account perturbative in
$\alpha_s(m_b)$ corrections. In two-loop approximations $m_b({\rm
pole})= \bar m_b(\bar m_b)(1+ 0.09 +0.05)$ \cite{ref.24}, where
the second and third numbers come from the $\alpha_s$ and
$\alpha_s^2$ corrections, respectively. In our calculations
$m_b({\rm pole})=4.83$~GeV is used, which corresponds to the
conventional current mass $\bar m_b(\bar m_b)\simeq 4.24$~GeV.

\item  $H_0$, as well as the mass $M(nl)$, does not contain an overall
additive (fitting) constant.

\item The string tension $\sigma=0.18$~GeV$^2$, used in the RSH,
cannot be considered a fitting parameter, because it is fixed
by the slope of the Regge trajectories for light mesons.

\item In the GE potential the asymptotic freedom behavior of the
vector strong coupling $\alpha_{\rm B}(r)$ is taken into account,
being expressed via the ``vector" QCD constant $\Lambda_{\rm B}$,
which is not a fitting parameter but defined by the conventional
$\Lambda_{\overline{MS}}$ according to the relation: $\Lambda_{\rm
B}(n_f=3)=1.4753~\Lambda_{\overline{MS}}(n_f=3)$ and $\Lambda_{\rm
B}(n_f=5)=1.3656~\Lambda_{\overline {MS}}(n_f=5)$ \cite{ref.25}. On the
other hand, the value of $\Lambda_{\overline{MS}} (n_f=5)$ is fixed by
the known value of $\alpha_s(M_Z)$ at the scale $M_Z=91.19$~GeV. Here
$\alpha_s(M_Z)=0.1191$ is used, which  in two-loop approximation
gives $\Lambda_{\overline {MS}}(n_f=5)=240$~MeV and correspondingly,
$\Lambda_{\rm B}(n_f=5)\simeq 330$~MeV.

\end{enumerate}

Thus our scheme of calculations appears to be very restrictive in
the case of bottomonium and only small variations of the fundamental
parameters are admissible. However, some uncertainty comes from
the value of the freezing constant, $\alpha_{\rm B}(r\rightarrow
\infty)\equiv \alpha_{\rm crit}$, which properties are discussed
in Ref.~\cite{ref.26}. Here we use the vector coupling in the range
$0.49\leq \alpha_{\rm crit}\leq 0.60$. Then for a given multiplet
$nl$ the centroid mass $M_{\rm cog}(nl)$ coincides with the eigenvalue
$M(nl)$ of the SSE:
\begin{equation}
 \left[2\sqrt{\vep^2 + m_b^2} + V_{\rm B}(r)\right]\varphi_{nl}
 =M(nl)\varphi_{nl} .
\label{eq.6}
\end{equation}
For this relativistic equation the NR limit and the so-called $einbein$
approximation may also be used and in both approximations a good
description of the bottomonium spectrum is obtained, even for the
higher states \cite{ref.20}. For most radiative decays (in
bottomonium) the m.e. like $I(mS|r|nP_J)$ and $I(nP_J|r|mS)$
differ only by $10-20\%$ in the NR and relativistic cases,
with the exception of the transitions $\Upsilon(3S)\rightarrow
\gamma+\chi_{bJ}(1P)$. In this case our calculations give
$\overline{I(3S|r|1P)}=0.007$~GeV$^{-1}$ in the NR case, being $\sim 3$
times smaller than $\overline{I(3S|r|1P)}=0.023$~GeV$^{-1}$ for the SSE
(here $\alpha_{\rm crit}=0.49$ was used). Notice that for a stronger GE
potential with $\alpha_{\rm crit}=0.60$ these spin-averaged m.e. appear
to be larger: $\overline{I(3S|r|1P)}=0.011$~GeV$^{-1}$ in the NR case
and 0.036~ GeV$^{-1}$ for the SSE.

Since the same static potential is used for the SSE as in the NR case,
such a difference between the m.e. may be explained by two factors: the
different asymptotic behavior of the w.f. of the SSE and Schr\"{o}dinger
equations, and also a smaller value of the w.f. at the origin for the
Schr\"{o}dinger equation as compared to that for the SSE. However,
it is known that the w.f. $R_{nS}(r)$, as well as the derivative
$R_{1P}'(r)$ for the $1P$ state, diverge near the origin for the SSE
(these divergences are discussed in details in Ref.~\cite{ref.16}) and
the calculated values of the w.f. (or its derivative) at the origin are
obtained with the use of a regularization procedure. This regularization
introduces a theoretical error, which is estimated to be $\la 10\%$.

In Table I we give the m.e. $I(3S|r|1P_J)$, calculated here for
the SSE and in the NR limit, together with their values from
second paper of Ref.~\cite{ref.13} (table 4.16) for the NR and the
relativistic variant RA, where a scalar confining potential, as in
our calculations, is used.

\begin{table}
\caption{The m.e. $I(3S|r|1P_J)$ (in GeV$^{-1}$) in the relativistic and
NR cases}
\begin{tabular}{lllll}
\hline
\hline
 Transition    &  ~  NR  &   ~ RA$^{a)}$&  ~ NR$^{b)}$ & ~SSE\\
\hline
     &   ~[13]~& ~[13]~ & ~this paper~& ~this paper~\\
\hline
~$\langle 3S|r|1P_2 \rangle$~ & ~0.016  & ~0.063 & ~ 0.047 & ~ 0.066\\
\hline
~$\langle 3S|r|1P_1 \rangle$~ & ~0.011  & ~0.063 & ~-0.033 & ~-0.020\\
\hline
~$\langle 3S|r|1P_0 \rangle$~ & ~0.004  & ~0.063 & ~-0.073 & ~-0.063\\
\hline
\hline
\end{tabular}

$^{a)}$Given numbers refer to the variant RA \cite{ref.13}, where a scalar
linear potential is used.\\
$^{b)} $ Given numbers refer to the NR limit of the SSE Eq.~(\ref{eq.6})
with the same potential $V_{\rm B}(r)$ and $\alpha_{\rm crit}=0.49$.
\end{table}
Comparison of the m.e. presented in Table I shows  that

\begin{enumerate}

\item In Ref.~\cite{ref.13} for the relativistic variant RA the
m.e. $I(3S|r|1P_J)$ is $\la 4$ times larger than in the NR case; a
similar result is obtained here for the spin-averaged m.e., where
$\overline{I(3S|r|1P)}=0.023~$GeV$^{-1}$ for the SSE and is an
$\sim 3$ times smaller value 0.007 GeV$^{-1}$ in its NR limit.

\item Corrections $\delta I_{\rm so}(J)= I(3S|r|1P_J)-
I\overline{(3S|r|1P)}$, due to the spin-orbit potential, have a
relatively large value, e.g. $\delta I_{\rm so}(J=2)= 0.043$~GeV$^{-1}$,
being almost two times larger than $\overline{I(3S|r|1P)}$ in the
spin-averaged case (see Eq.~(\ref{eq.9}) below).

\item In Ref.~\cite{ref.13} the splittings between the m.e.
$I(3S|r|1P_J)$ with different $J$ are much smaller than in our
calculations.

\item In the spin-orbit potential we take the strong coupling
$\alpha_{\rm so}(\mu)=0.38$, which is close to the value $\alpha_{\rm
so}(\mu(2P))$ used for the $\chi_{bJ}(2P)$ states (this value was
extracted in Ref.~\cite{ref.23} from the experimental masses of the
members of the $\chi_{bJ}(2P)$ multiplet). Our calculations here
show that the nondiagonal m.e., like $\lan nP|V_{\rm so}|mP \ran~(n\neq m,
n=1,2,3)$, are of the same order or have even larger values than the
diagonal m.e. $\lan 2P|V_{\rm so}(r)| 2P\ran$.

\end{enumerate}

The calculated $E1$ transition rates are presented in Table II
together with their values from Ref.~\cite{ref.13}; they
correspond to the m.e. from Table I.

\begin{table}
\caption{The partial widths $\Gamma(\Upsilon(3S)\rightarrow
\gamma+\chi_b(1P_J))$ (in eV)}
\begin{tabular}{llcccll}
\hline
\hline
 ~Transition ~ & ~ $E_{\gamma}$~ & ~ RA ~ & ~ NR ~& ~ SSE ~& ~ exp. ~ & ~ exp.\\
\hline
 & ~ (MeV)~ & ~ [13] ~& ~ this paper ~& ~ this paper ~& ~ CLEO [6] ~& ~ BaBar [8] ~\\
\hline
 ~$\Gamma_2(\Upsilon(3S)\rightarrow \gamma+\chi_b(1P_2))$~ & ~ 433.5~ &~ 195 ~& 108 ~& 213~& ~ $157\pm 30$ ~& ~ $216\pm 25$\\
\hline
 ~$\Gamma_1(\Upsilon(3S)\rightarrow \gamma+\chi_b(1P_1))$~ & ~ 452.1~ &~ 134~& 36 ~& 13~& ~~ $33\pm 10$~& $ ~~ < 22$ \\
\hline
 ~$\Gamma_0(\Upsilon(3S)\rightarrow \gamma+\chi_b(1P_0))$~ & ~ 483.9~ &~~ 54 ~& 72 ~& 54~& ~~ $61\pm 23$~& ~~ $55\pm 10$\\
\hline
\hline
\end{tabular}
\end{table}

In the relativistic case our transition rates appear to be very close
to those from the BaBar data \cite{ref.9}. Even in the NR case, due to
large spin-orbit corrections, the calculated partial widths do not
contradict the CLEO data \cite{ref.8}.

We make some remarks on the contribution $\delta I_{\rm so}$ to
the m.e. $I(3S|r|1P_J)$ from the spin-orbit potential,
$\hat{V}_{\rm so}(r)=\veL \cdot \veS~ V_{\rm so}(r)$, for which
the splittings $a_{\rm so}(nP|1P)=\lan nP|V_{\rm
so}|1P\ran,~(n=2,3)$ are taken as for the one-gluon exchange
interaction, i.e., neglecting the second order corrections in
$\alpha_s(\mu)$ (it may be shown that the second order corrections
are negative and small, $\sim -0.7$~MeV). In this approximation we
find
\be
 a_{\rm so} (nP,1P) = \frac{1}{2\omega^2_b} {\{4\alpha_{\rm so} \lan
 r^{-3}\ran_{nP,1P}- \sigma \lan r^{-1}\ran_{nP,1P}\}},
\label{eq.7}
\ee
where we take $\alpha_{\rm so}=0.38$, which provided a good description
of the fine-structure splittings for the $\chi_b(2P_J)$ multiplet. To
determine the corrections to the w.f. of the $\chi_{bJ}(1P)$ states,
the potential $\hat {V}_{\rm so}$ is considered as a perturbation and
the following mass differences between the centroid masses are used:
\be
 M_{\rm cog} (2P) - M_{\rm cog}(1P) = 360~{\rm MeV}, ~M_{\rm cog} (3P) -
 M_{\rm cog}(1P) = 640~{\rm MeV}. \label{eq.8}\ee
Notice that the correction from the $3P$ state is not small, while the
value of the centroid mass $M(3P)$, $M(\chi_b(3P))\simeq 10.54$~GeV,
is taken from the recent ATLAS experiment \cite{ref.27}.

For the SSE, the splittings $a_{\rm so}(2P,1P)=12$~MeV and $a_{\rm
so}(3P,1P)= 10.2$~MeV, were calculated and in the NR limit their
values are $\sim 10$\% smaller. Then the nondiagonal m.e.
$I(3S|r|1P_J)$ with the ``spin-orbit" corrections can be presented
(in GeV$^{-1}$) as
\begin{eqnarray}
 I(3S|r|1P_J) & = & \overline{I(3S|r|1P)} + \delta I_{\rm so}(J),
\nonumber \\
 \delta I_{\rm so}(J) & = & 0.033~ \xi_J~ \overline{I(3S|r|2P)} + 0.016~ \xi_J
 ~\overline{I(3S|r|3P)},
\label{eq.9}
\end{eqnarray}
where $\xi_J=-2, -1, +1$ for $J=0,1,2$ and
$\overline{I(3S|r|1P)}=0.023$~GeV$^{-1}$ for the SSE (relativistic
case) and 0.007 GeV$^{-1}$ in the NR limit. To obtain the m.e.
presented in Table I, we use also the spin-averaged nondiagonal
m.e.: $\overline{I(3S|r|2P)}=-2.54$~GeV$^{-1}$ and
$\overline{I(3S|r|3P)}=2.64$~GeV$^{-1}$.

\section{Conclusions}

For the $E1$ radiative transitions, $\Upsilon(3S)\rightarrow
\gamma+\chi_b(1P_J)~(J=0,1,2)$, the spin-averaged m.e.
$\overline{I(3S|r|1P_J)}$ are shown to be small, as it was predicted
in a number of studies before.

However, due to spin-orbit effects the w.f. of the $1\,{}^3P_1$
state is mixed with the $2P,3P$ states, for which the m.e.
$\overline{I(3S|r|2P)}$ and $\overline{I(3S|r|3P)}$ are large and
have different signs. Such a mixing is important, although the
spin-orbit splittings themselves are not large and their typical
values are $\sim 10-12$~MeV. Due to this mixing, a strong
cancellation takes place in the m.e. $I(3S|r|1P_1)$, which gives
rise to a suppression of the transition rate for the radiative decay
$\Upsilon (3S)\rightarrow \gamma+\chi_b(1\,{}^3P_1)$.

The following partial widths are predicted:
$\Gamma_J(\Upsilon(3S)\rightarrow \gamma+\chi_b(1\,{}^3P_J))=213$~eV,
13~eV, and 54~eV for $J=2,1,0$, which are in good agreement with the
BaBar data, $\Gamma_2({\rm exp}.)=216\pm 25$~eV and $\Gamma_0({\rm
exp}.)=55\pm 10$~eV \cite{ref.9}. Also for $J=1$ the calculated partial
width $\Gamma_1=13$~eV satisfies the upper limit, $\Gamma_1 < 22$~eV,
obtained in the BaBar experiment. More precise measurements of the
transition rate for $\Upsilon(3S)\rightarrow \gamma+\chi_{b1}(1P)$
could give additional restrictions on the spin-orbit effects in
radiative decays.

We predict the following ratio of the partial widths:,
$r_{1,0}=\frac{\Gamma_1}{\Gamma_0}=0.24$, which should be
considered as an important feature of the transition rates where
spin-orbit dynamics dominates.

\begin{acknowledgments}

The authors are grateful to J.~L.~Rosner and Yu.~A.~Simonov for
useful remarks and suggestions.

\end{acknowledgments}


\begin{thebibliography}{99}

\bibitem{ref.1}
G.~Bonvicini et al. (CLEO Collab.), Phys. Rev. D {\bf 70}, 032001
(2004).

\bibitem{ref.2}
P.~del A.~Sanchez et al. (BaBar Collab), Phys. Rev. Lett. {\bf
93}, 162002 (2004).

\bibitem{ref.3}
B.~Aubert et al. (BaBar Collab.), Phys. Rev. Lett.  {\bf 101},
071801 (2008).

\bibitem{ref.4}
B.~Aubert et al. (BaBar Collab.), Phys. Rev. Lett. {\bf 103},
161801  (2009).

\bibitem{ref.5}
G.~Bonvicini et al. (CLEO Collab.), Phys. Rev.  {\bf 81}, 031104
(R) (2010).

\bibitem{ref.6}
M.~Artuso et al. (CLEO Collab.), Phys. Rev. Lett.  {\bf 94}, 032001
 (2005).

\bibitem{ref.7}
D.~M.~Asner et al. (CLEO Collab.), Phys. Rev. D {\bf 78}, 091103
(2008).

\bibitem{ref.8}
M.~Kornicer et al. (CLEO Collab.), Phys. Rev. D {\bf 83}, 054003
(2011); arXiv: 1012.0589 (2010) [hep-ex].

\bibitem{ref.9}
J.~P.~Lees et al. (BaBar Collab.), arXiv:1104.5254 (2011)
[hep-ex].

\bibitem{ref.10}
W.~Kwong and J.~L.~Rosner, Phys. Rev. D {\bf 38}, 279 (1988).

\bibitem{ref.11}
N.~Brambilla et al., Eur. Phys. J. C {\bf 71}, 1534 (2011).

\bibitem{ref.12}
E.~Eichten, S.~Godfrey, H.~Mahlke, and J.~L.~Rosner, Rev. Mod.
Phys. {\bf 80}, 1161(2008).

\bibitem{ref.13}
D.~Ebert, R.~N.~Faustov, and V.~O.~Galkin, Phys. Rev. D {\bf 67},
014027 (2003); N.~Brambilla et al., arXiv: hep-ph/0412158 (2004).


\bibitem{ref.14}
J.~L.~Rosner, arXiv:1107.1273 (2011) [hep-ph] and references
therein.

\bibitem{ref.15}
H.~Grotch, D.~A.~Owen, and K.~J.~Sebastian, Phys. Rev. D {\bf
30}, 1924 (1984); S.~N.~Gupta, S.~F.~Radford, and W.~W.~Repko,
Phys. Rev. D {\bf 30}, 2424 (1984).

\bibitem{ref.16}
S.~Jacobs, M.~G.~Olsson, and C.~Suchyta, Phys. Rev. D {\bf 33},
3338 (1986).

\bibitem{ref.17}
A.~Grant and J.~L.~Rosner, Phys. Rev. D {\bf 46}, 3862 (1992).

\bibitem{ref.18}
A.~M.~Badalian, B.~L.~G.~Bakker, and I.~V.~Danilkin, Phys. Atom.
Nucl. {\bf 73}, 138 (2010); arXiv:0903.3643 (2009) [hep-ph].

\bibitem{ref.19}
A.~Yu.~Dubin, A.~B.~Kaidalov, and Yu.~.A.~ Simonov, Phys. Atom.
Nucl. {\bf 56}, 1745 (1993); hep-ph/9311344; Phys. Lett. B {\bf
323}, 41 (1994); Yu.A.~Simonov, hep-ph/9911237 (1999).

\bibitem{ref.20}
A.~M.~Badalian, B.~L.~G.~Bakker, and I.~V.~Danilkin, Phys. Rev. D
{\bf 81}, 071502 (2010), Erratum-ibid. D {\bf 81}, 099902 (2010);
ibid. D {\bf 79}, 037505 (2009); A.~M.~Badalian, A.~I.~Veselov,
and B.~L.~G.~Bakker, Phys. Rev. D {\bf 70}, 016007 (2004).

\bibitem{ref.21}
D.~P.~Stanley and D.~Robson, Phys. Rev. D {\bf 21}, 3180 (1980);\\
W.~Lucha, F.~F.~Schoberl, and D.~Gromes, Phys. Rept. {\bf 200},
127 (1991) and references therein.

\bibitem{ref.22}
S.~Godfrey and N.~Isgur, Phys. Rev. D {\bf 32}, 189 (1985).

\bibitem{ref.23}
A.~M.~Badalian and  B.~L.~G.~Bakker, Phys. Rev. D {\bf 62}, 094031
(2000).

\bibitem{ref.24}

K.~Nakamura et al. (Particle Data Group) J. Phys. G  {\bf 37},
075021 (2010).

\bibitem{ref.25}
M.~Peter,  Phys. Rev. Lett. {\bf 78}, 602 (1997); Y.~Schroder,
Phys. Lett. B {\bf 447}, 321 (1999).

\bibitem{ref.26}
Yu.~A.~Simonov, arXiv:1011.5386 (2010) [hep-ph]; A.~M.~Badalian,
and A.~I.~Veselov, Phys. Atom. Nucl. {\bf 68}, 582 (2005);
A.~M.~Badalian and D.~S.~ Kuzmenko, Phys. Rev. D {\bf 65}, 016004
(2002).

\bibitem{ref.27}
G.~Aad et al. (ATLAS Collab.), arXiv:1112.5154 (2011) [hep-ex]

\end{thebibliography}
\end{document}